\documentclass[twocolumn,nofootinbib,preprintnumbers]{revtex4-1}
\usepackage{pslatex}
\usepackage[pdftex]{graphicx}

\usepackage{psfrag}
\usepackage{epsfig}
\usepackage{color}
\usepackage{cancel}
\usepackage{amsmath}
\usepackage{enumitem,xspace}

\def\refe@jnl#1{{#1}}
\def\aj{\refe@jnl{Astron.~J.}}
\def\araa{\refe@jnl{Annu.~Rev.~Astron.~Astrophys.}}
\def\apj{\refe@jnl{Astrophys.~J.}}
\def\apjl{\refe@jnl{Astrophys.~J.~Lett.}}
\def\aap{\refe@jnl{Astron.~Astrophys.}}
\def\mnras{\refe@jnl{Mon.~Not.~R.~Astron.~Soc.}}
\def\prd{\refe@jnl{Phys.~Rev.~D}}
\def\fcp{\refe@jnl{Fund.~Cos.~Phys.}}
\def\physrep{\refe@jnl{Phys.~Rep.}}
\def\physlett{\refe@jnl{Phys.~Lett.}}

\def\invisible#1{  }

\def\lsim{\mathrel{\lower4pt\hbox{$\sim$}} 
\hskip-9.5pt\raise1.6pt\hbox{$<$}\;} 
 \def\gsim{\mathrel{\lower4pt\hbox{$\sim$}} 
\hskip-9.5pt\raise1.6pt\hbox{$>$}\;}

\definecolor{orange}{rgb}{1,0.5,0}

\begin{document}
\preprint{ULB-TH/13-16}

\title{Charged lepton flavor violation and the origin of neutrino masses}

\author{Thomas Hambye}
\email{thambye@ulb.ac.be; Review talk given at the "1$^{st}$ Conference on Charged Lepton Flavor Violation", May 6-8 2013, Lecce, Italy}
\affiliation{Service de Physique Th\'eorique\\
 Universit\'e Libre de Bruxelles\\ 
 Boulevard du Triomphe, CP225, 1050 Brussels, Belgium}


\begin{abstract}
The neutrino oscillations 
imply 
that charged lepton flavor violation (CLFV) processes do exist.
Even if the associated rates are in general expected very suppressed, it turns out that this is not always necessarily the case.
In the framework of the three basic seesaw models, 
we review the possibilities of having observable rates and thus, in this way, of distinguishing these possible neutrino mass origins. 
\end{abstract}

 \maketitle

\section{Introduction}
\label{}

As abundantly discussed during this workshop, the search for charged lepton flavor violation is expected to know in the next few years a real step forward.  This concerns a long series of processes with a $\mu-e$, $\tau-\mu$ or $\tau-e$ transition.
For $\mu-e$ processes an improvement of up to 4 to 6 orders of magnitude could be expected, in particular  for the $\mu\rightarrow e e e$ \cite{Berger:2011xj} decay and $\mu$ to $e$ conversion in atomic nuclei \cite{Hungerford:2009zz}-\cite{Kurup:2011zza}. Important improvements are also expected for the $\mu\rightarrow e \gamma$ decay \cite{MEG2}. On the theory side such transitions 
could be induced by a large variety of beyond the SM physics models. This new physics could lead to observable rates even if the associated energy scale is in some cases as large as few thousands of TeV. At the moment it is not clear which type of new physics will manifest itself at such energies. 
In this talk 
we will review the possibility to get observable rates from the physics associated to the neutrino masses, the only beyond the Standard Model (SM) physics which has been established so far at the laboratory level (gravitation put apart). It is well known that neutrino oscillations guarantee non-vanishing CLFV but at a very suppressed level. Beyond this experimentally unreachable contribution we will review the possibilities  that the seesaw states, that are generally expected to be at the origin of the neutrino masses, could induce rates that are not suppressed by the smallness of these masses. This offers an opportunity to probe better the yet unknown neutrino mass origin.

\section{The 3 seesaw frameworks}

If neutrinos oscillate, CLFV processes can be induced through the following lepton chain: $l_i { \xrightarrow{W}} \nu_{Li}\rightarrow \nu_{Lj} { \xrightarrow{W}} l_j$, with the $W$ index referring to a vertex involving a $W$ boson.
No matter the neutrinos are of Dirac or Majorana type, the $\nu_{Li}\rightarrow \nu_{Lj} $ transition  requires two neutrino mass insertions. As a result the CLFV rates are suppressed by 4-power of the neutrino masses. For example for $\mu\rightarrow e \gamma$ one gets
\begin{equation}
Br(\mu\rightarrow e \gamma)=\frac{3}{32}\frac{\alpha}{\pi}\frac{\sum_{i=e,\mu,\tau}|{\cal M}_{\nu e i}{\cal M}^\dagger_{\nu i \mu}|^2}{m_W^4}\sim 10^{-52}
\label{CLFVmnu}
\end{equation}
where the last equality results from assuming typically ${\cal M}_{\nu_{ij}}\sim 0.1~\hbox{eV}$, with ${\cal M}_\nu$ the neutrino mass matrix.
The corresponding rate is extremely small and hopeless for experimentalists. For the Majorana case, and unlike the Dirac case, one expects nevertheless other contributions to the CLFV rates. Majorana masses, as in the favorite seesaw way of generating neutrino masses, require 
the existence of new states with their own masses, that can induce CLFV transitions in ways not suppressed by the neutrino masses.

\begin{figure}[t]
\centering
\includegraphics[height=1.58cm]{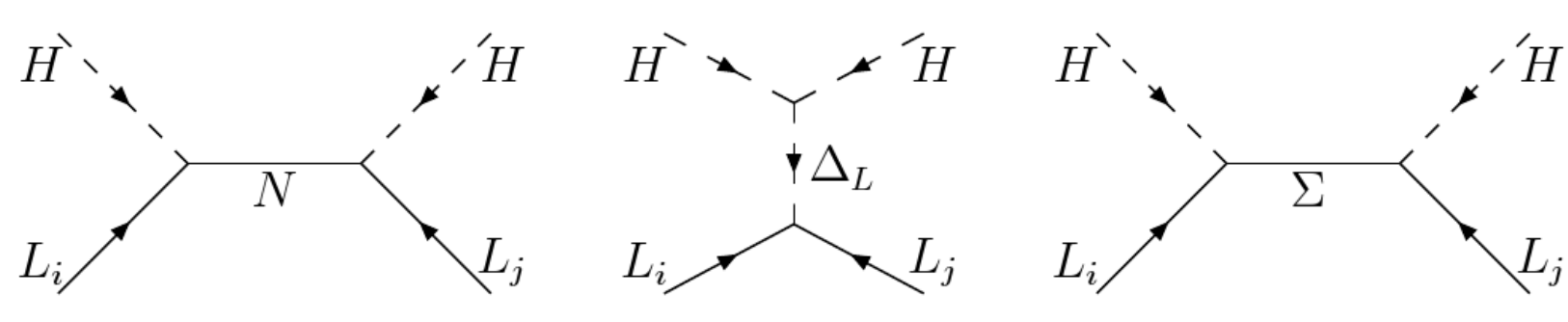}\\
\caption{The 3 basic seesaw diagrams which can induce naturally small neutrino masses.}
\label{fig1}
\end{figure}

There are 3 ways of inducing $\Delta L=2$ Majorana neutrino masses from the tree level exchange of a heavy particle: from the exchange of right-handed neutrinos $N_i$ (seesaw of type-I), of a scalar triplet $\Delta_L$ (type-II) and of fermion triplets $\Sigma_i$ (type-III). This is illustrated in Fig.~\ref{fig1}. In the first and third case lepton number is broken from the coexistence of Majorana masses for the heavy states and Yukawa interactions. The latter couple the heavy states to the Standard Model scalar doublet $H=(H^+,H^0)^T$ and to a lepton doublet $L=(\nu_\alpha,l^-_\alpha)^T$
\begin{eqnarray}
{\cal L}&\owns&\, -\frac{1}{2} m_{N_i} \overline{N}_i  N_i^c 
    -Y_{N_{ij}} \tilde{\phi}^\dagger \overline{N}_i  L_j +h.c.\\
 {\cal L}&\owns& \,  -\frac{1}{2} m_{\Sigma_i}Tr [\overline{\Sigma}   \Sigma^c] 
                  -\sqrt{2}Y_{\Sigma_{ij}} \tilde{\phi}^\dagger \overline{\Sigma}_i  L_j +h.c. 
\end{eqnarray}               
with 
\begin{equation}
\Sigma=
\left(
\begin{array}{ cc}
   \Sigma^0/\sqrt{2}  &   \Sigma^+ \\
     \Sigma^- &  -\Sigma^0/\sqrt{2} 
\end{array}
\right)\,,
\end{equation}
and $\tilde{H}=i\tau_2 H^\ast$. The neutrino mass matrix one gets in these cases is ${\cal M}_\nu^N=-\frac{v^2}{2} Y_N^T\frac{1}{m_N} Y_N$  and                
${\cal M}_\nu^\Sigma=-\frac{v^2}{2} Y_\Sigma^T\frac{1}{m_\Sigma} Y_\Sigma$, with  $v=246$~GeV. As for the type-II case, lepton number is broken from the fact that the scalar triplet couples to both a lepton doublet pair and a scalar doublet pair
\begin{equation}
{\cal L}\owns\, -m^2_\Delta Tr[\Delta_L^\dagger \Delta_L]     -L^TY_\Delta C i \tau_2\Delta_LL+\mu_\Delta \tilde{H}^Ti \tau_2 \Delta_L \tilde{H}          
\end{equation}
with 
\begin{equation}
\Delta_L=
\left(
\begin{array}{ cc}
   \delta^+/\sqrt{2}  &   \delta^{++} \\
     \delta^0 &  -\delta^+/\sqrt{2} 
\end{array}
\right)\,,
\end{equation}
giving ${\cal M}_\nu^\Delta=Y_\Delta\mu_\Delta^*\frac{v^2}{m^2_\Delta}$.

Neutrino masses require electroweak symmetry breaking. In all seesaw models they involve two powers of the electroweak symmetry breaking scale, $v=\sqrt{2} \langle H^0\rangle\simeq 246$~GeV. As can be seen in Fig.~\ref{fig1}, in all cases the neutrino mass generation proceeds in the same way. An effective interaction between 2 $L$ and to 2 $H$ is induced, which, once we replace both neutral scalar fields by $v$, leads to the neutrino masses above. The $LLHH$ effective interaction induced has a single possible form, the dimension 5 Weinberg operator form, ${\cal L}_{eff}^{d=5}=\frac{c^{d=5}_{\alpha\beta}}{\Lambda} (\overline{L^c}_\alpha \tilde{H}^\ast)(\tilde{H}^\dagger L_\beta)$, with 
$ {\cal M}_{\nu_{\alpha\beta}}=-\frac{v^2}{2}c^{d=5}_{\alpha\beta}/\Lambda$. The $\Lambda$ scale can be identified as the overall scale where the seesaw states lie, $m_N$, $m_\Delta$ or $m_\Sigma$. 
Clearly, since the effective interaction induced is the same in all cases, and since any of the 3 seesaw models above could give any possible neutrino mass matrix (i.e.~any $c^{d=5}_{\alpha\beta}$ coefficients for this interaction), the knowledge of the neutrino mass matrix is not sufficient
to be able to distinguish between the 3 seesaw models. One would need additional information from other kinds of experiments. CLFV could provide this information, provided a series of conditions is satisfied.

\section{Large CLFV rates in seesaw models}

To illustrate the fact that seesaw states can induce CLFV rates that are not suppressed by the smallness of the neutrino masses, the type-II seesaw case is parti\-cularly clear. The seesaw scalar triplet does not induce only the lepton number violating dimension 5 effective interaction. It also induces a lepton number conserving dimension 6 effective interaction which induces CLFV. From the exchange of a scalar triplet and 2 $Y_\Delta$ Yukawa couplings, a four lepton effective interaction is induced, 
\begin{equation}
{\cal L}_{eff}^{d=6}\owns c_{\alpha\beta\delta\gamma}^{d=6} (\overline{L}_\beta \gamma_\mu L_\delta)(\overline{L}_\alpha \gamma^\mu L_\gamma)
\label{O6II}
\end{equation}
with $c^{d=6}_{\alpha\beta\delta\gamma} =(1/2m^2_\Delta) Y^\dagger_{\Delta_{\alpha \beta}} Y_{\Delta_{\delta\gamma}}$. Clearly such a process can induce a $\mu\rightarrow eee$ or $\tau \rightarrow 3l$ decay at tree level and $l\rightarrow l'\gamma$ or $\mu\rightarrow e$ conversion at one loop. In this way we get in particular \cite{Barger:1982cy}
\begin{equation}
Br(\mu\rightarrow eee)=|c^{d=6}_{\mu eee}|^2/G_F^2=|Y_{\Delta e \mu}|^2 |Y_{\Delta ee}|^2/(4 m^4_\Delta G_F^2)\,.
\label{mueeeII}
\end{equation}
The CLFV rates in this case are quadratic in the dim-6 coefficient rather than quartic in the dim-5 one (as in Eq.~(\ref{CLFVmnu}), i.e.~are proportional to $(Y_\Delta/m_\Delta)^4$ rather than to $(Y_\Delta/m_\Delta)^4\cdot (\mu_\Delta/m_\Delta)^4$).
The dim-6 contribution could be many orders of magnitude larger than the dim-5 one, especially if $m_\Delta$ is low.
For example with $m_\Delta=1$~TeV, and $\mu_\Delta/m_\Delta\sim Y_\Delta\sim 10^{-6}$ (so that $m_\nu\sim 0.1$~eV), one gets $Br(\mu\rightarrow eee)\sim 10^{-27}$. But $\mu_\Delta/m_\Delta$ has no reason to be similar to $Y_\Delta$. If furthermore $Y_\Delta>> \mu_\Delta/m_\Delta$, one gets further enhanced rates. For example, still with $m_\Delta=1$~TeV, and with $Y_\Delta\sim 10^{-2.5}$ (and $\mu_\Delta/m_\Delta\sim 10^{-9}$ so that $m_\nu\sim 0.1$~eV) one saturates the present experimental sensitivity for this rate.
In other words, if the seesaw scale is relatively low, and if the smallness of the neutrino masses is due to a large part to the smallness of  the $\mu_\Delta/m_\Delta$ parameter, one can get large rates. 
Of course to have a so low seesaw scale goes against the usual seesaw expectation that the smallness of the neutrino masses is due mostly to 
a very large seesaw scale, but nothing forbids such a possibility. To consider a small $\mu_\Delta$ parameter is technically natural because it is protected by a symmetry. In the $\mu_\Delta\rightarrow 0$ limit lepton number is conserved.

In the framework of the type-I and type-III seesaw models, a similar situation is also possible even if an approximately $L$ conserving setup doesn't show up in a so straightforward way. 
From the exchange of a right-handed neutrino or fermion triplet and 2 Yukawa interactions, and without $N$ or $\Sigma$ Majorana mass insertion in their propagator, the lepton number conserving dimension 6 effective interactions that are 
induced are \cite{Broncano:2002rw,Abada:2007ux}
\begin{eqnarray}
{\cal L}_{eff}^{d=6}&=&c_{\alpha\beta}^{d=6} \overline{L}_\alpha \tilde{\phi}i\partial\hspace{-1.7mm}\slash(\tilde{\phi}^\dagger L_\beta)\\
{\cal L}_{eff}^{d=6}&=&c_{\alpha\beta}^{d=6} \overline{L}_\alpha \tau^a \tilde{\phi}iD\hspace{-2mm}\slash(\tilde{\phi}^\dagger \tau^aL_\beta)
\end{eqnarray}
with $c^{d=6}_{\alpha\beta}=(Y_N^\dagger \frac{1}{M_N^2} Y_N)_{\alpha\beta}$ and $c^{d=6}_{\alpha \beta}=(Y_\Sigma^\dagger \frac{1}{M_\Sigma^2} Y_\Sigma)_{\alpha\beta}$ respectively.
Clearly such effective interactions also induce CLFV rates that have no reason to be suppressed in the same way than the dimension 5 contribution. The former contribution is quadratic in the L conserving dim-6 coefficients, whereas the latter one is quartic in the L violating dim-5 one  (i.e.~involve 8 rather than 4 Yukawa couplings). 
As a result, for example in the type-I scenario, we get
\begin{equation}
Br(\mu\rightarrow e \gamma)=\frac{3}{8}\frac{\alpha_{em}}{\pi} \sum_i\frac{|Y_{N_{ie}}Y^\dagger_{N_{i\mu}}|^2}{m_{N_i}^4} v^4
\label{muegammalownaive}
\end{equation}
For $m_{N_i}$ close to the GUT scale this still results in very suppressed rates. However for low values of $m_{N_i}$ the rates we get are much larger. Using the typical seesaw expectation $Y_N^2\sim m_\nu m_N/v^2$,
with $m_N$ the typical seesaw scale, for $m_N \sim 100$~GeV one gets $Br(\mu\rightarrow e \gamma)\sim 10^{-26}$. Therefore, as for the type-II case above, low scale seesaw generically gives much larger rates than in Eq.~(\ref{muegammalownaive}), but still too small to be reached experimentally by the next generation of experiments. However the typical seesaw expectation used here to relate Yukawa couplings to neutrino masses has no reason to be necessarily valid. The $c^{d=6}_{\alpha\beta}$ coefficients involve the Yukawa couplings in a $Y Y^\dagger$ $L$ conserving combination, whereas the $c^{d=5}$ coefficients involve them in a $Y^T Y$ $L$ violating one.
Since both combinations differ on the basis of a symmetry they have no reason to be related in such a simple way. 
In particular lepton number is not necessarily broken as soon as a fermion seesaw states have masses and Yukawa interactions. It turns out that the fermion seesaw states can have masses and Yukawa interactions without breaking lepton number, see e.g.~Refs.~\cite{Abada:2007ux}-\cite{Blanchet:2009kk}. The most simple example consists in considering only 2 right-handed neutrinos (or fermion triplets), and to assign a $U(1)_L$ lepton number symmetry under which
$N_1$ has $L=1$, $N_2$ has $L=-1$, see in particular Ref.~\cite{Gavela:2009cd}.
This $U(1)_L$ symmetry allows only a Majorana mass term coupling both $N's$, $\propto m_{N} \overline{N}_1 N_2^c +h.c.$, and Yukawa couplings only for $N_1$, $\propto 
    -Y_{N_j} \tilde{\phi}^\dagger \overline{N}_1  L_j$. As a result $N_1$ and $N_2$ mixes maximally to form two degenerate mass eigenstates with mass $m_N$, that both have Yukawa couplings. In this symmetry limit, since lepton number is conserved, neutrinos are massless, no matter how large the $Y_{N_j}$ couplings are, so that dim-6 coefficients, and hence CLFV rates, can be large independently of the size of the neutrino masses. This simply requires large Yukawa couplings for $N_1$ and a relatively low $m_N$ scale. 
In order that neutrino masses are induced, the $U(1)_L$ symmetry must be broken by small parameters, Yukawa couplings for $N_2$ and diagonal right-handed mass terms (in $\overline{N_1}N_1^c$ and $\overline{N_2}N_2^c$, which gives a $N$ mass splitting), leading to neutrino masses proportional to these L breaking parameters. Of course such a scenario is not anymore the minimal seesaw in the sense that it requires to assume an additional approximate $U(1)_L$ symmetry setup but it doesn't require any additional states or interactions beyond the ones of the minimal seesaw model.\footnote{Note that these scenarios in some cases can be minimal in the sense of minimal flavor violation, i.e.~from the knowledge of the full flavor structure of the neutrino mass matrix, they allow to know the full flavor structure of the dim-6 coefficients \cite{Gavela:2009cd}. This means that they predict the full flavor structure of the CLFV rates. Together with the determination of the seesaw scale $m_N$, which can be done as explained below, this allows a full reconstruction of the seesaw lagrangian.} 
We will now explain how such a possibility could be singled out from CLFV processes. Note that these setups predict a quasi-degenerate spectrum of right-handed neutrinos, since the mass splitting breaks $L$. This will play an important role below.

\section{Type-I seesaw CLFV predictions}

\begin{figure}[t!]
    \centering
        \includegraphics[width=0.365\textwidth]{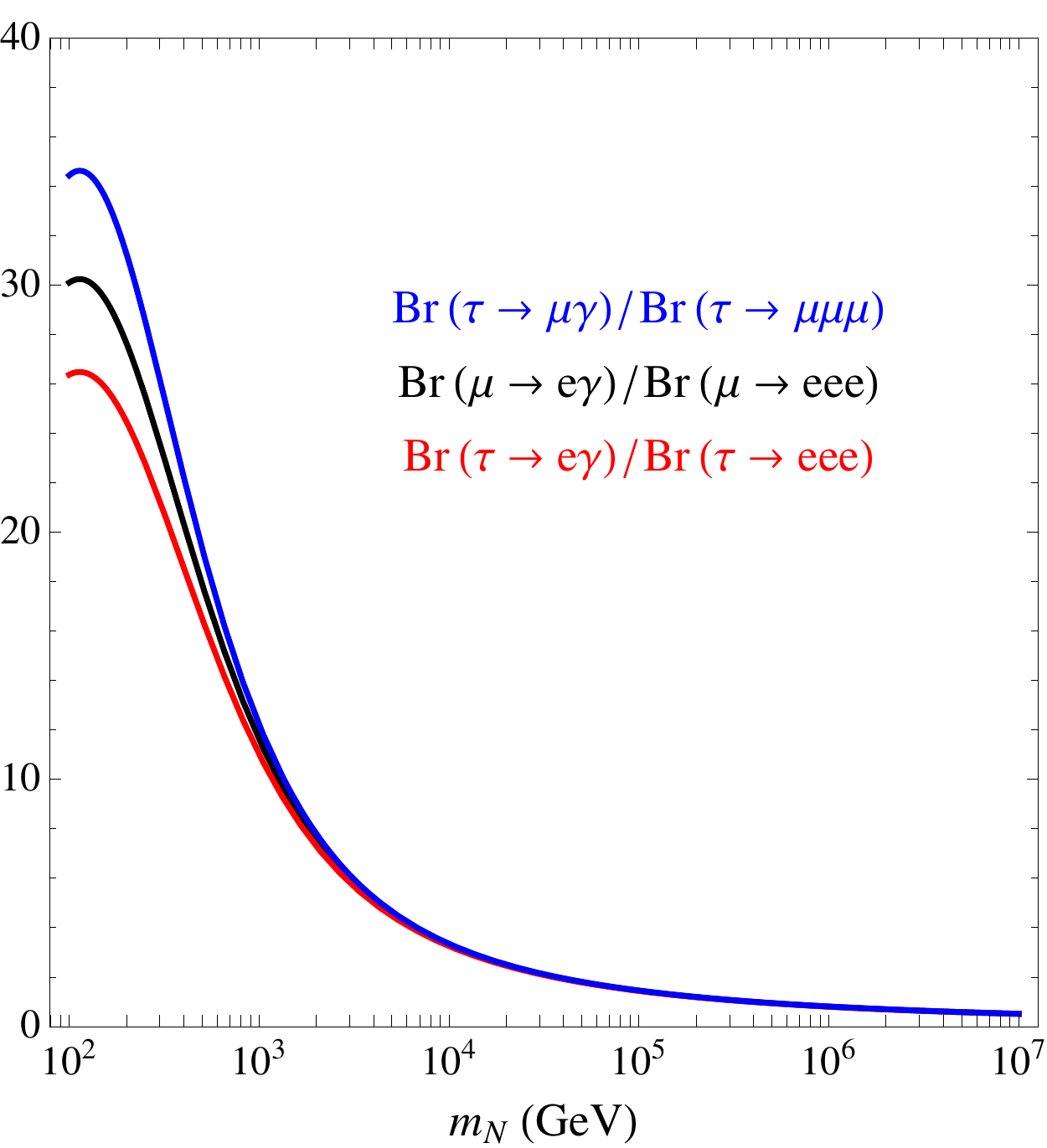} 
        \caption{
        $R^{l \rightarrow l' \gamma}_{\mu\to l'l'l'}= Br(l \rightarrow l' \gamma)/Br(\mu \rightarrow l'l'l')$ as a  function of $m_N$, from Refs.~\cite{Chu:2011jg,Alonso:2012ji}.}
        \label{Plot1}
\end{figure}
In the type-I seesaw model all CLFV processes are necessarily induced at the loop level because flavor violation in this model occurs only at the level of the neutral leptons, requiring at least one internal $W$ boson inside the loop diagram.
The $\mu\rightarrow e \gamma$ rate has been calculated already very long ago, see Refs.~\cite{Minkowski:1977sc}-\cite{Cheng:1980tp}. For the $\mu\rightarrow eee$ rate, see Ref.~\cite{Ilakovac:1994kj}. The $\mu\rightarrow e$ conversion rate has been calculated by a series of references with different results. This recently motivated us to redo carefully this calculation, see Ref.~\cite{Alonso:2012ji} and Refs.~therein. 
As pointed out in Refs.~\cite{Chu:2011jg} and \cite{Alonso:2012ji}, to test the possible seesaw origin of one or several CLFV processes, that could be observed in the future, the most promising way is to consider ratios of two CLFV processes involving a same $l\rightarrow l'$ transition.
This statement is supported by the following considerations.
First, all process rates involving a same flavor transition have the same general form
\begin{equation}
T_{l\rightarrow l'}= \sum_{N_i}\frac{|Y_{N_{il'}}Y^\dagger_{N_{il}}|}{m_{N_i}^4} \cdot [c+c'log(m_{N_i}^2/m_W^2)]^2
\label{generalrateform}
\end{equation}
where $c$ and $c'$ are dimension four numerical factors that depend on $m_l$, $m_{W,Z,h}$ and on the process considered (neglecting the mass of $l'$). 
$T_{l\rightarrow l'}$ holds here for any CLFV processes, for example $Br(\mu\rightarrow e \gamma)$, $Br(\mu \rightarrow eee)$ or $\mu\rightarrow e$ conversion rate in atomic nuclei, $R^N_{\mu\rightarrow e}$ (defined in Ref.~\cite{Alonso:2012ji} for example).
Second, as we have seen, the setups that can lead in a natural way to observable CLFV rates are based on an approximate $L$ symmetry setup that implies a quasi-degenerate spectrum of right-handed neutrinos.
As a result, in Eq.~(\ref{generalrateform}), the right-handed neutrino mass splittings can be neglected and the dependence of the rates in the Yukawa couplings factorizes out. Therefore, in the ratio of two same flavor transition rates, the Yukawa coupling dependence cancels out and we are left with a function which depends only on a single unknown parameter, the overall $m_N$ scale
\begin{equation}
\frac{T^{(1)}_{l\rightarrow l'}}{T^{(2)}_{l\rightarrow l'}}=\frac{(c_{1}+c'_{1}log(m_{N_i}^2/m_W^2))^2}{(c_{2}+c'_{2}log(m_{N_i}^2/m_W^2))^2
}=fct(m_N)
\end{equation}
 This allows several possibilities of tests we will now discuss.

\begin{figure}[t]
    \centering
        \includegraphics[width=0.45\textwidth]{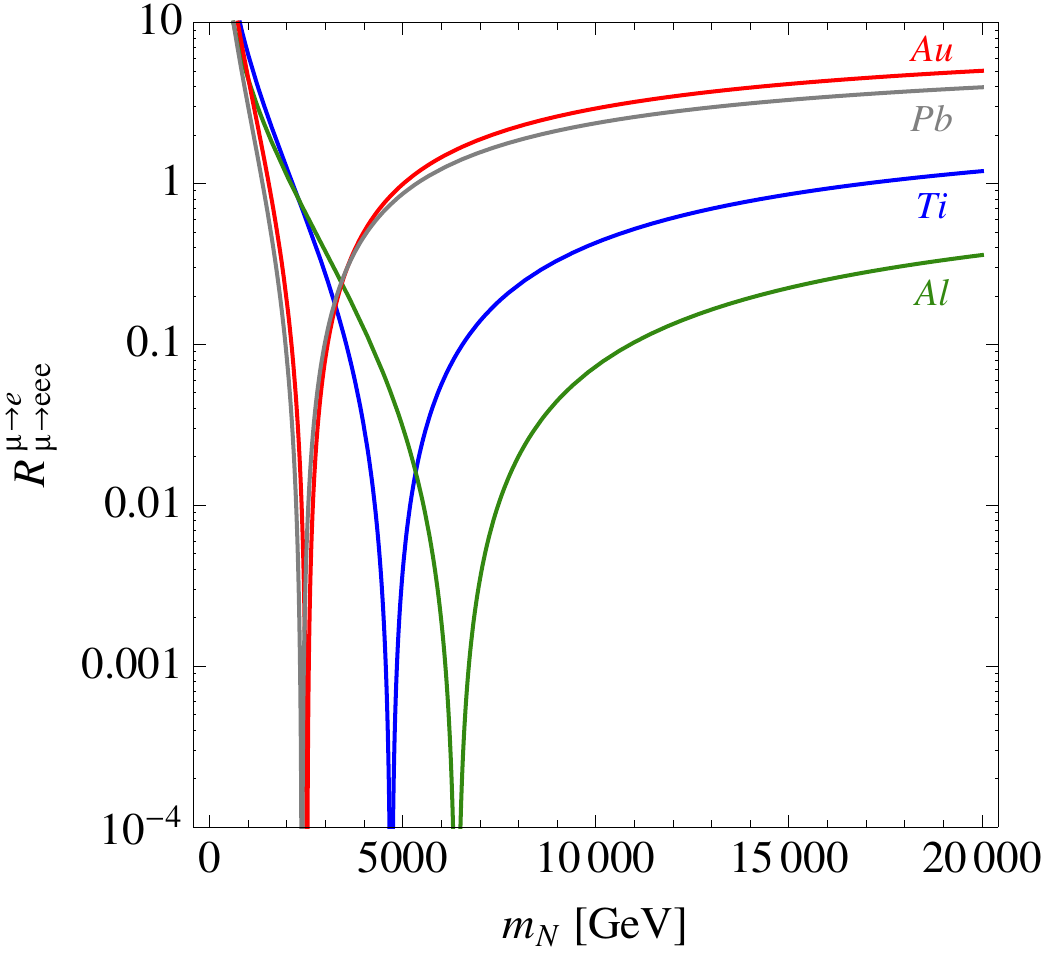} 
        \caption{
        $R^{\mu \rightarrow e}_{\mu\to eee}= R^N_{\mu\rightarrow e}/Br(\mu\to eee)$ as a function of the right-handed neutrino mass scale $m_N$, for $\mu \rightarrow e$ conversion in various nuclei, from Ref.~\cite{Alonso:2012ji}.}
        \label{Plot2}
\end{figure}

In Figs.~\ref{Plot1} and \ref{Plot2} are plotted the various ratios one gets for the various CLFV processes.
First of all, for the $Br(\mu\rightarrow e\gamma)$ to $Br(\mu\rightarrow eee)$ ratio, one observes \cite{Chu:2011jg,Alonso:2012ji} a monotonous function that is always larger than $\sim2.5$ (for values of $m_N$ that do give an observable rate), which means that the measurement of this ratio would basically allow for a determination of $m_N$, or for an exclusion of the scenario if a ratio smaller than $\sim 2.5$ is found.
In fact the observation of a single rate could already be sufficient to exclude the model, if together with the experimental upper bound on the other one, it leads to a ratio incompatible with the expectations of Fig.~\ref{Plot1}.   
A same discussion holds for the other ratios \cite{Alonso:2012ji} given in Fig.~\ref{Plot2}, except for the fact that the $R^N_{\mu\rightarrow e}$ conversion rates (which depend crucially on the nuclei considered) turn out to vanish for a particular value of $m_N$. Consequently the ratios are not monotonous functions of $m_N$. The value of $m_N$ which gives a vanishing $R_{\mu\rightarrow e}^N$ depends on the nuclei considered. This illustrates how important it would be to search for $\mu\rightarrow e$ conversion, not only with one nuclei, but with several of them. The degeneracy in $m_N$ that a single ratio can display could be lifted up by the measurement of another ratio.
The reason why these rates vanish for a particular value of $m_N$ can be traced back to the fact that the up quark and down quark contributions have different signs in the amplitude, depending on their different charges and weak isospins (and display different $m_N$ dependences), see Ref.~\cite{Alonso:2012ji}.

\begin{figure}[t!]
    \centering
        \includegraphics[width=0.4\textwidth]{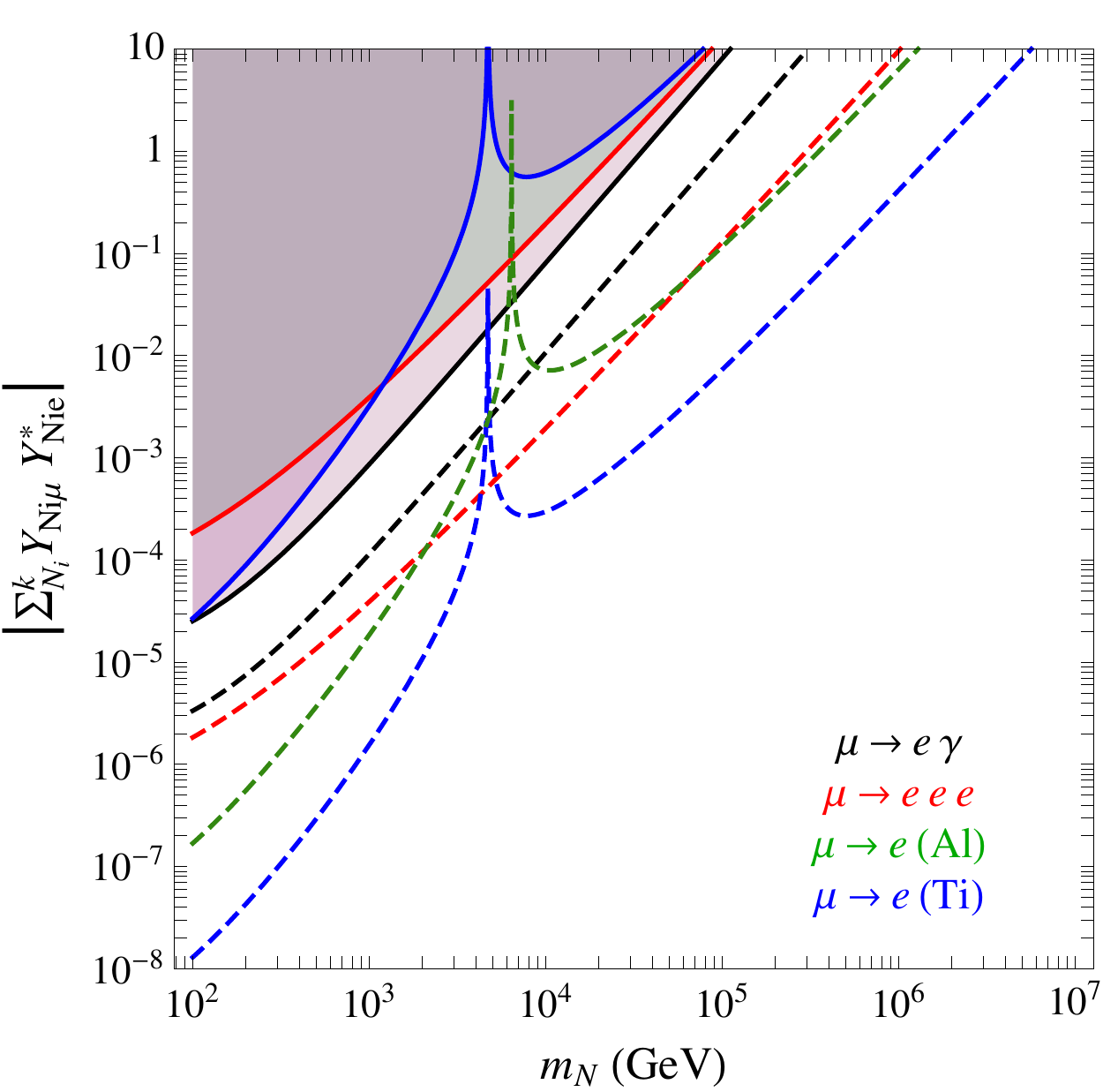} 
        \caption{From Ref.~\cite{Alonso:2012ji}, present bounds and future sensitivity  on $|\sum_{i}^k Y_{\Sigma_{i\mu}}  Y_{\Sigma^*_{ie}}|$ 
for scenarios characterized by one right-handed neutrino mass scale. The solid lines are obtained from present experimental upper bounds: from $R^{Ti}_{\mu\rightarrow e} < 4.3\cdot 10^{-12}$ \cite{Dohmen:1993mp} and $Br(\mu\rightarrow e \gamma)<5.7\cdot 10^{-13}$~\cite{Adam:2013mnn}, $Br(\mu\rightarrow eee)<10^{-12}$~\cite{Bellgardt:1987du}. The dashed lines are obtained from the expected experimental sensitivities: from $R^{Ti}_{\mu\rightarrow e} \lesssim 10^{-18}$ \cite{Hungerford:2009zz,Cui:2009zz},  $R^{Al}_{\mu\rightarrow e} \lesssim 10^{-16}$ \cite{Hungerford:2009zz}-\cite{Kurup:2011zza} and $Br(\mu\rightarrow e \gamma)<10^{-14}$~\cite{MEG2}, $Br(\mu\rightarrow eee)<10^{-16}$~\cite{Berger:2011xj}.}
        \label{Plot3}
\end{figure}

Fig.~\ref{Plot3} shows the lower bounds resulting for the Yukawa couplings, if  the various rates are required to be large enough to be observed in planned experiments. It also shows the upper bounds which hold today on these quantities from the non-observation of these processes. This  figure illustrates well the impact of future $\mu \rightarrow e$ conversion measurements/bounds, as they will become increasingly dominant in exploring  flavor physics in the $\mu-e$ charged lepton sector. Values of the Yukawa couplings as low as $10^{-2}$, $10^{-3}$ and $10^{-4}$ could be probed,  for $m_N=100$~TeV, 1~TeV and 100~GeV, respectively, with Titanium experiments being the most sensitive.
If the Yukawa couplings are for example of order unity, the bounds of Fig.~\ref{Plot3} can be rephrased as upper bounds on the $m_N$ scale:
\begin{eqnarray}
m_N&\lesssim&2000\,\hbox{TeV}\cdot \Big(\frac{10^{-18}}{R^{Ti}_{\mu\rightarrow e}}\Big)^{\frac{1}{4}}\cdot a^{1/2}\,,\nonumber\\
m_N&\lesssim&300\,\hbox{TeV}\cdot \Big(\frac{10^{-16}}{R^{Al}_{\mu\rightarrow e}}\Big)^{\frac{1}{4}}\cdot a^{1/2}\,,\nonumber\\
m_N&\lesssim&100\,\hbox{TeV}\cdot \Big(\frac{10^{-14}}{Br(\mu\rightarrow e \gamma)}\Big)^{\frac{1}{4}}\cdot a^{1/2}\,,\nonumber\\
m_N&\lesssim&300\,\hbox{TeV}\cdot \Big(\frac{10^{-16}}{Br(\mu\rightarrow e ee)}\Big)^{\frac{1}{4}}\cdot a^{1/2}\,\nonumber.
\end{eqnarray}
with $a= |\sum_{N_i} Y_{N_{ie}}Y^\dagger_{N_{i\mu}}|$.
Overall, this exercise shows that future experiments may in principle probe the type-I seesaw model beyond the $\sim 1000$~TeV scale, and that $\mu\rightarrow e$ conversion future experiments could become with time the most sensitive ones.

Finally in Fig.~\ref{Plot1} can also be found the predictions the type-I seesaw scenario give for the ratio of 2 same flavor transition processes involving the $\tau$ lepton, $\tau\rightarrow \mu$ and $\tau \rightarrow e$ transitions.
The current and future expected limits on such processes do not allow to access $m_N$ scale as large as from the $\mu\rightarrow e$ processes, but neutrino data still allows that the $\tau$ processes would be boosted with respect to the $\mu\rightarrow e$ ones (depending on the mass hierarchy considered). For specific Yukawa configurations of this kind, compare for example the minimal models considered in Ref.~\cite{Gavela:2009cd,Chu:2011jg}.

\section{Type-III seesaw CLFV predictions}

There is a crucial difference between the type-I and type-III seesaw models for what concerns CLFV processes. While in the type-I case there is flavor mixing only at the level of the neutral leptons, for the type-III case there is flavor mixing directly at the level of the charged leptons. For instance a $\mu^-$-$e^-$ transition can proceed directly through charged lepton/charged $\Sigma$ mixing, i.e. through the $\mu^-{ \xrightarrow{Y}} \Sigma^- { \xrightarrow{Y}} e^-$ chain, with the $Y$ index referring to a vertex involving a Yukawa coupling (i.e.~the insertion of a SM scalar boson or its vev).
As a result, if for the type-I case all processes necessarily occurs at the loop level, for the type-III case the $l\rightarrow 3l$ and $\mu\rightarrow e$ conversion process in atomic nuclei proceed at tree level (just attach for example a $Z$ boson on the fermionic chain above and couple it on the other side to 2 leptons or 2 quarks). Only $\mu\rightarrow e \gamma$ still has to proceed at loop level because the QED coupling remains flavor diagonal in the charged fermion mass eigenstate basis (unlike Z couplings, see Refs.~\cite{Abada:2007ux,Abada:2008ea}). The calculation of all corresponding rates has recently been performed in Refs.~\cite{Abada:2008ea}. Obviously the tree level processes do not induce logarithmic terms and have the simple general form
\begin{equation}
T_{l\rightarrow l'}= \sum_{\Sigma_i}\frac{|Y_{\Sigma_{il'}}Y^\dagger_{\Sigma_{il}}|}{m_{\Sigma_i}^4} \cdot c
\label{generalrateformtypeIII}
\end{equation}
with $c$ a constant which depends only on $m_l$ and $m_{W,Z}$ (neglecting $m_{l'}$). As a result the ratio of a same flavor transition process are predicted to a fixed value! The measurement of such a ratio could then easily rule-out the type-III scenario as possible explanation of these processes if another value is obtained. Alternatively it would provide a strong indication for it if the right ratio was observed. As for the $\mu\rightarrow e \gamma$ rate, it turns out that it also doesn't display any logarithmic term and has the general form of Eq.~(\ref{generalrateformtypeIII}), leading to fixed ratios too, when compared with the other rates. Alltogether we get the neat predictions \cite{Abada:2008ea}
\begin{eqnarray}
Br(\mu\rightarrow e \gamma)&=&1.3\cdot 10^{-3}\cdot Br(\mu\rightarrow eee)\nonumber\\
&=&3.1\cdot 10^{-4}\cdot R^{Ti}_{\mu\rightarrow e}\nonumber
\end{eqnarray}
Larger $\mu\rightarrow eee$ and $\mu\rightarrow e$ conversion rates, as compared to the $\mu\rightarrow e \gamma$ rate, are characteristic of this model. In the majority of other beyond the standard model scenarios that can lead to observable CLFV rates, one finds $Br(\mu\rightarrow eee)<Br(\mu\rightarrow e \gamma)$.

As for the sensitivity to the overall seesaw scale $m_\Sigma$, it is even more impressive than for the type-I seesaw case, in particular for the processes which proceed at tree level, 
\begin{eqnarray}
m_\Sigma&\lesssim&20000\,\hbox{TeV}\cdot \Big(\frac{10^{-18}}{R^{Ti}_{\mu\rightarrow e}}\Big)^{\frac{1}{4}}\cdot a^{1/2}\,,\nonumber\\
m_\Sigma&\lesssim&3000\,\hbox{TeV}\cdot \Big(\frac{10^{-16}}{R^{Al}_{\mu\rightarrow e}}\Big)^{\frac{1}{4}}\cdot a^{1/2}\,,\nonumber\\
m_\Sigma&\lesssim&100\,\hbox{TeV}\cdot \Big(\frac{10^{-14}}{Br(\mu\rightarrow e \gamma)}\Big)^{\frac{1}{4}}\cdot a^{1/2}\,,\nonumber\\
m_\Sigma&\lesssim&1600\,\hbox{TeV}\cdot \Big(\frac{10^{-16}}{Br(\mu\rightarrow e ee)}\Big)^{\frac{1}{4}}\cdot a^{1/2}\,\nonumber.
\end{eqnarray}
with $a= \sum_{\Sigma_i}|Y_{\Sigma_{ie}}Y^\dagger_{\Sigma_{i\mu}}|$.
 
Finally for the $\tau$ decay processes one gets the predictions
\begin{eqnarray}
Br(\tau\rightarrow \mu \gamma)&=&1.3\cdot 10^{-3}\cdot Br(\tau\rightarrow \mu\mu\mu)\nonumber\\
&=&2.1\cdot 10^{-3}\cdot Br(\tau\rightarrow \mu^-e^-e^+)\nonumber\\
Br(\tau\rightarrow e \gamma)&=&1.3\cdot 10^{-3}\cdot Br(\tau\rightarrow eee)\nonumber\\
&=&2.1\cdot 10^{-3}\cdot Br(\tau\rightarrow e^- \mu^+\mu^-)\nonumber
\end{eqnarray}

\section{Type-II seesaw CLFV predictions}

\begin{figure}[t!]
    \centering
        \includegraphics[width=0.365\textwidth]{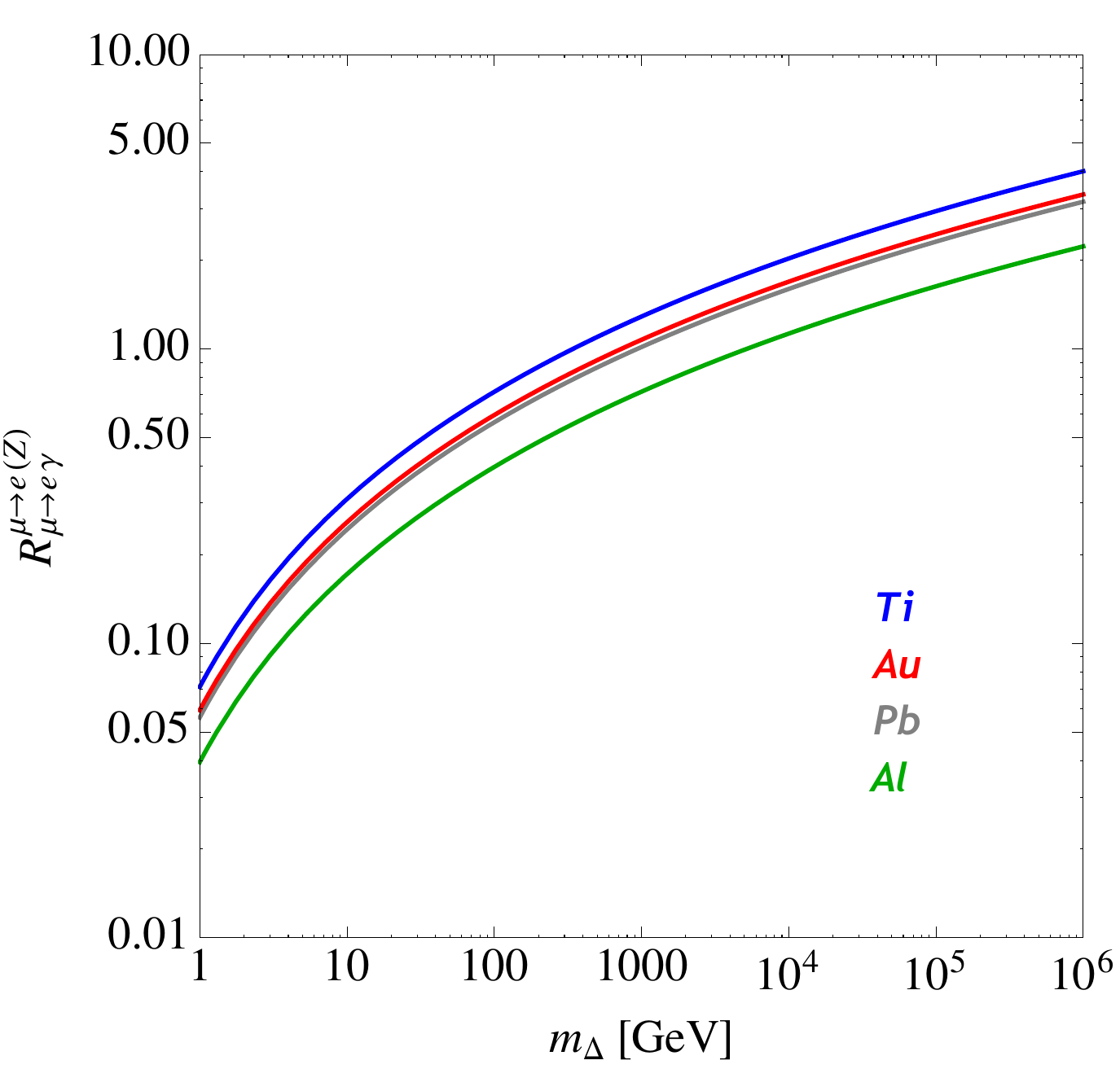} 
        \caption{$R_{\mu\rightarrow e}^{N}/Br(\mu\rightarrow e\gamma)$ for various nuclei, as a  function of $m_\Delta$.}
        \label{Plot5}
\end{figure}
Each seesaw model presents a different pattern for what concerns the various CLFV process. As we saw above, for the type-I model all processes are loop processes, and for the type-III case they are tree level processes, except $l\rightarrow l'\gamma$. The type-II model presents an intermediate situation.  On the one hand from the exchange of a scalar triplet and 2 $Y_\Delta$ interactions one gets the 4 lepton interactions of Eq.~(\ref{O6II}), hence $l\rightarrow 3l$ at tree level, Eq.~(\ref{mueeeII}). On the other hand the other processes can only be induced at one loop from the exchange of a scalar triplet between 4 leptons too, but necessarily contracting 2 of the leptons in a loop.
As a result, the $l \rightarrow 3l$ rates are expected to be comparatively the largest one.
Compared to the fermion seesaw cases, another important difference is that, by taking the ratio of two same flavor transition rates, one does not always get a function which depends only on the seesaw state mass. For instance the $l\rightarrow 3l$ rates  do not involve the same Yukawa couplings than the other processes. For example $Br(\mu\rightarrow eee)$ involves the product of $Y_{\Delta_{e \mu}}$ and $Y^\dagger_{\Delta_{ee}}$, Eq.~(\ref{mueeeII}), whereas $\mu\rightarrow e\gamma$ and $\mu\rightarrow e$ conversion processes involve a sum over $l=e,\mu,\tau$ of $Y_{\Delta_{l \mu}} \cdot Y^\dagger_{\Delta_{le}}$. 
This means that a ratio involving $\mu\rightarrow eee$ and another processes depends on the Yukawa coupling configuration considered, but still, we would like to point out here that any ratio involving any processes but $\mu\rightarrow eee$ depends only on $m_\Delta$ (for $m_\Delta>>> m_{e,\mu\tau}$). Taking the~type-II  $\mu\rightarrow e \gamma$ rate in e.g.~Refs.~\cite{typeIImuegamma,Abada:2007ux,Dinh:2012bp} and $\mu\rightarrow e$ conversion rate in e.g.~Ref.~\cite{Dinh:2012bp}, in Fig.~\ref{Plot5} we give the values of these ratios. These ratios are monotonous functions of $m_\Delta$.\footnote{We thank M.~Dhen for having produced this plot and for discussions about it.} In the same way as for the type-I case, the measurement of any of these ratios would therefore gives us the value of $m_\Delta$ if these processes are due to the type-II seesaw interactions (or exclude this model as their origins). The observation of $\mu\rightarrow e\gamma$ should nevertheless comes first.
This can be seen from the sensitivity on $m_\Delta$ we get for the various processes 
\begin{eqnarray}
m_\Delta\hspace{-1mm}&<& \hspace{-1mm}2200\, \hbox{TeV}\cdot \sqrt{|Y_{\Delta_{\mu e}}||Y_{\Delta_{ee}}|}\cdot \Big(\frac{10^{-16}}{Br(\mu\rightarrow e ee)}\Big)^{\frac{1}{4}}\,\,\,\nonumber\\
m_\Delta\hspace{-1mm}&<& \hspace{-1mm}15\, \hbox{TeV}\cdot \sqrt{|Y_{\Delta_{\tau\mu}}||Y_{\Delta_{\mu\mu}}|}\cdot \Big(\frac{10^{-8}}{Br(\tau\rightarrow \mu\mu\mu)}\Big)^{\frac{1}{4}}\nonumber\\
m_\Delta\hspace{-1mm}&<& \hspace{-1mm}8\, \hbox{TeV}\cdot \sqrt{|Y_{\Delta_{\tau e}}||Y_{\Delta_{ee}}|}\cdot \Big(\frac{10^{-8}}{Br(\tau\rightarrow e ee)}\Big)^{\frac{1}{4}}\nonumber
\end{eqnarray}
and
\begin{eqnarray}
m_\Delta\hspace{-1mm}&<&\hspace{-1mm} 70\, \hbox{TeV}\cdot |\sum_l Y_{\Delta_{\mu l}} Y^\dagger_{\Delta_{le}}|^{\frac{1}{2}}\cdot \Big(\frac{10^{-14}}{Br(\mu\rightarrow e \gamma)}\Big)^{\frac{1}{4}}\nonumber\\
m_\Delta\hspace{-1mm}&<& \hspace{-1mm}600\, \hbox{TeV}\cdot |\sum_l Y_{\Delta_{\mu l}} Y^\dagger_{\Delta_{le}}|^{\frac{1}{2}} \cdot \Big(\frac{10^{-18}}{R^{Ti}_{\mu\rightarrow e}}\Big)^{\frac{1}{4}}\,.\nonumber
\end{eqnarray}

\section{Summary}

In conclusion neutrino oscillations guarantee that CLFV processes do exist.
If the neutrino masses are of the Dirac type the effect is nevertheless extremely suppressed ($Br(\mu\rightarrow e\gamma)\sim 10^{-52}$). If they are of the Majorana type, and if the seesaw states at their origin lies close to the GUT scale, they are also very suppressed. If instead the seesaw states lie around the 100 GeV-1 TeV scale they are expected hugely larger but still generically too small for the next generation of CLFV experiments ($Br(\mu\rightarrow e\gamma)\sim 10^{-26/-30}$). However without adding any new seesaw states or interactions, it turns out that there exist special configurations of the couplings that can lead to rates saturating the present CLFV upper bounds. These are technically natural because can be justified on the basis of a symmetry (approximate lepton number conservation). For such cases each seesaw setup comes with a quite different CLFV pattern. In particular for the type-III case the ratio of two same flavor transition processes is totally predicted (no matter the values of the fermion triplet masses) and offers a very clear possibility of test. Similarly in the type-I case, for the case where the right-handed neutrino would have a quasi-degenerate mass spectrum (which is precisely what happens in the setups that lead to observable rates), these ratios depend only on the common seesaw mass scale. This also offers clear possibility of tests.
For the type-II case, the $\mu\rightarrow eee$ process should clearly be the first observed rate in the future, whereas for the other ones $\mu\rightarrow e$ conversion will become gradually the most sensitive probe.

\vspace{5mm}
\section*{Acknowledgement}
\vspace{-3mm}

It is a pleasure to thank A. Abada, R. Alonso, C. Biggio, F. Bonnet, X. Chu, M. Dhen, B. Gavela, D. Hernandez and P. Hernandez with whom the work presented here has been done. We thank also useful discussions with L. Calibbi. This work was supported by the FNRS-FRS, the IISN and the Belgian Science Policy, IAP VII/37.

\end{document}